# Development of compact gas treatment system using secondary emission electron gun

M. Watanabe, Y. Wang, A. Okino, K. C. Ko[*], E. Hotta

*Department of Energy Sciences, Tokyo Institute of Technology, Yokohama, JAPAN*
*\* Department of Electrical and Computer Engineering, Hanyang University, Seoul, KOREA*
*E-mail: watanabe@es.titech.ac.jp*

**Abstract**

It is well known that the non-thermal plasma processes using electrical discharge or electron beam are effective for the environmental pollutant removal. Especially, the electron beam can efficiently remove pollutant, because a lot of radicals which are useful to remove pollutant can be easily produced by high-energy electrons.

We have developed a compact 100kV secondary emission electron gun to apply $NO_X$ removal. The device offers several inherent advantages such as compact in size, wide and uniform electron beam. Besides, the device offers good capability in high repetition rate pulsed operation with easy control compared with glow discharge or field emission control cathode guns.

In present study, the $NO_X$ removal characteristics have been studied under the increased gun voltage, varied pulsed electron beam parameters such as current density and pulse width as well as gas flow rate. The experimental results indicate a better $NO_X$ removal efficiency comparing to other high-energy electron beam and electrical discharge processing.

**1. Introduction**

Exhaust gas treatment is regarded as a stringent research for the environment protection because the exhaust gas, such as $NO_X$, $SO_X$, $CO_2$, has been the main causes of the acid rain and the greenhouse effect. $NO_X$ and $SO_X$ are exhausted by the power plants in large quantities, and are the direct cause of the acid rain. Some methods have been developed for the $NO_X$ treatment. Various types of plasma devices have been employed for the environmental applications. Exhaust gas treatment technologies by plasma process have been actively investigated and found to be significantly effective. In all the cases, the non-thermal plasma is used to generate high-energy electrons that collide with background molecules to produce radicals, ions and secondary electrons, which in turn, decompose the gaseous pollutants. Exhaust gas treatment technologies by non-thermal plasma process include pulse corona process, barrier discharge process, DC corona, electron beam and so on. According to these processes, electron beam and electrical discharge are the two methods employed in various ways to decompose the gaseous pollutants at atmospheric pressure. In either case, the high-energy electrons serve to produce the radicals, ions and secondary electrons that in turn decompose the gaseous pollutants.

In our experiments, the secondary electron beam produced by the secondary emission electron gun[1] is used to irradiate and decompose the $NO_X$ gas. In present study, the NOx removal characteristics have been studied under the increased gun voltage, varied pulsed electron beam parameters such as current density and pulse width as well as gas flow rate. The experimental results indicate a better NO removal efficiency comparing to other high-energy electron beam and electrical discharge processing.

## 2. Experimental Setup

Figure 1 shows the schematic diagram of the experimental device, including a wire ion plasma source (WIPS), a secondary emission electron gun (SEEG) and a gas treatment chamber. The WIPS is a thin wire glow discharge device, which is used as a high-density plasma source as well as an ion source. The positive helium ions ($He^+$) extracted from the WIPS are accelerated in vacuum towards a negatively biased stainless steel cathode. When such ions collide on the cathode surface, a number of secondary electrons are generated due to a kinetic emission process. The emitted secondary electrons are then accelerated towards an electron window and form a very wide and uniform electron beam. Here, uniquely the electron window is kept on the side orthogonal to the ion extraction window. This configuration is referred as side-extraction-type (SET). The conventional design is referred to as vertical incident type[2], in which the emitted electrons are accelerated again towards the discharge region and collided with particles in WIPS before passing to the electron window. The main advantage of this kind of configuration over the conventional vertical incident type is that it avoids the interaction between the electron beam and the WIPS. The experiment is carried out using a pulsed WIPS discharge and continuous negative acceleration.

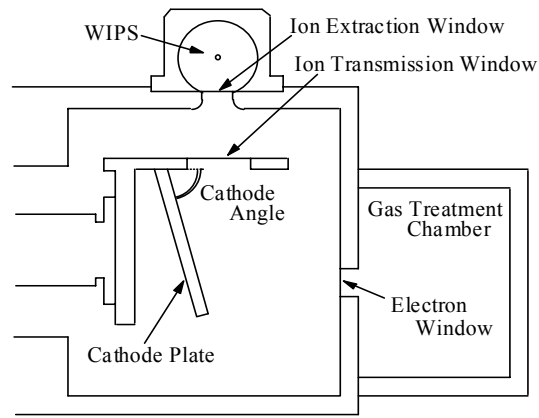

Fig.1 Schematic diagram of side-extraction-type secondary emission electron gun.

The volumes of the gas treatment chamber is 3.5 L. Considering that the penetration depth of the 100 keV electron beam in 1 atmospheric pressure $N_2$ is about 110 mm, the gas chamber with the height of 110 mm has been designed and used to act as the $NO_X$ gas treatment chamber in order to achieve the adequate irradiation.

## 3. Results and Discussion

The mechanism of $NO_X$ removal is considered as a complex process, which involves a large number of possible chemistry reactions. The electron beam energy absorbed in background gas ($N_2$, NO) induces N and O species, which subsequently react with other molecular. The major decomposition reactions are given as follows:

$$e + N_2 \rightarrow e + N + N$$
$$e + O_2 \rightarrow e + O + O$$
$$N + NO \rightarrow N_2 + O$$
$$NO + O + M \rightarrow NO_2 + M \quad \text{(M represents other particles)}$$
$$NO_2 + O \rightarrow NO + O_2$$

The presence of N and O radicals leads to some intense reactions, by which the $NO_X$ can be reduced. The reduction of NO gas is dominated by the reaction: $N + NO \rightarrow N_2 + O$, and by a process of complex reactions, NO is decomposed to $N_2$ and $O_2$. The performance of the treatment is achieved by investigate the removal ratio of $NO_X$ (%) and removal efficiency of $NO_X$ (nmol/J), which are defined as the following equations:

$$\text{Removal ratio (\%)} = \frac{\text{Removed NO}_X \text{ (ppm)}}{\text{Initial concentration (ppm)}} \times 100$$

$$\text{Removal efficiency (nmol/J)} = \frac{\text{Removed NO}_X \text{ (ppm)} \times \text{Gas flow rate (L/s)} \times 10^{-3}}{24.4 \text{ (L/mol)} \times \text{Electron energy (J/shot)} \times \text{Pulse repetition (shot/s)}}$$

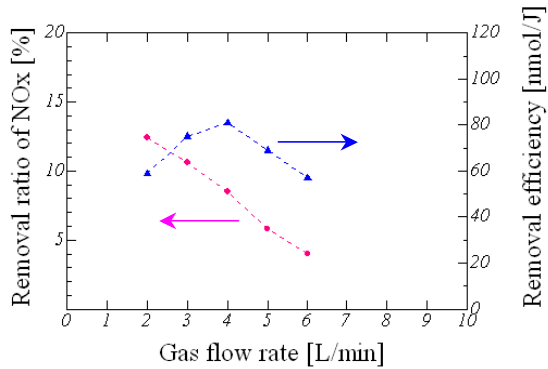

Fig.2 Measurements of $NO_X$ removal ratio and removal efficiency for various gas flow rates.

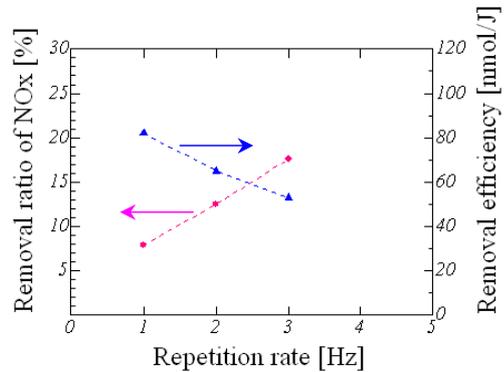

Fig.3 Measurements of NOx removal ratio and removal efficiency for various pulse repetition rates.

Figure 2 shows the dependence of $NO_X$ removal ratio and removal efficiency as a function of gas flow rate at a cathode voltage of 90 kV and a pulse repetition of 2 Hz. The removal ratio decreases with a increase of gas flow rate. This is because the energy of electron beam used for the decomposition $NO_X$ per unit of volume, per unit of time decreases in inverse proportion to gas flow rate. On the other hand the removal efficiency meets the maximum at the gas flow rate of 4 L/min. Figure 3 shows the dependence of $NO_X$ removal ratio and removal efficiency as a function of pulse repetition at a cathode voltage of 90 kV and a gas flow rate of 2 L/min. The removal ratio increases with an increase of pulse repetition, whereas the removal efficiency decreases with increase of pulse repetition. Figure 4 shows the dependence of $NO_X$ removal ratio as a function of cathode voltage at a pulse repetition of 1 Hz and a gas flow rate of 2 L/min. As shown in the Fig. 4, the electron beam with the energy over 20 keV can transmit the electron window, however the energy of electron beam for the effective decomposition is above 30 keV. And the removal ratio increases sharply over the cathode voltage of 70 kV. Figure 5 shows the dependence of $NO_X$ removal efficiency as a function of cathode voltage at a pulse repetition of 1 Hz and a gas flow rate of 2 L/min. The removal efficiency increases in proportion to cathode voltage. And the maximum removal efficiency of 107 nmol/J was obtained. It is found that the removal efficiency of SEEG is less than the highest removal efficiency at the first shot in pulsed electron beam of 160 keV (256 nmol/J) or continuous electron beam of 120 keV (310 nmol/J) [3,4]. Also it is found that the removal efficiency of SEEG is about 2.5 times better than that of the pulsed corona reactor (43 nmol/J) and that of the dielectric barrier discharge reactor (42 nmol/J) [5].

## 4. Summary

The electron beam generated by SEEG has the characteristics as follows: kinetic energy; $\leq$ 100 keV, current; ~2.2 A, current density; ~31.2 mA/cm$^2$, pulse width; ~10 μs and the electron beam energy per pulse; 2.2 J. The pulsed electron beam for the $NO_X$ dissociation is injected into the gas chamber filled with 1 atmospheric pressure $N_2$ with an initial concentration of 250 ppm of NO. The $NO_X$ removal experiments have been investigated in the condition that the flue gas is flowing through the gas chamber, which is different from the conventional study of non-flowing condition of flue gas. The removal ratio and removal efficiency were investigated in terms of gas flow rate, pulse repetition and cathode voltage respectively. And a maximum removal efficiency of 107 nmol/J was obtained. The further experiments are expected to be more efficient by increasing cathode voltage and pulse repetition.

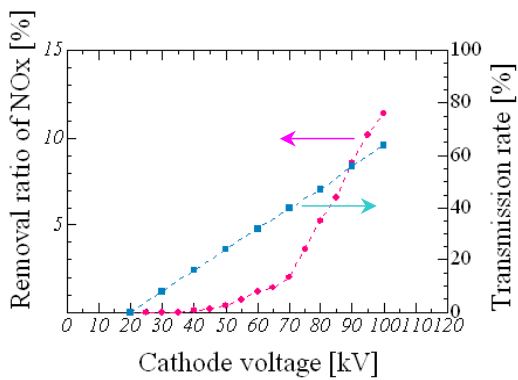

Fig.4 Measurements of $NO_X$ removal ratio for various cathode voltages.

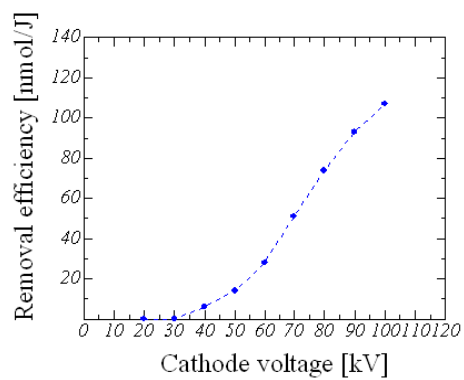

Fig.5 Measurements of NOx removal efficiency for various cathode voltages.

**References**

1. P. R. Chalise, Y. Wang, K. A. Mustafa, M. Watanabe, Y. Hayashi, A. Okino and E. Hotta, "NOx Treatment Using Low-Energy Secondary Emission Electron Gun", *IEEE Trans. on Plasma Science*, Vol. 32. No. 3, pp. 1392-1399, 2004.
2. H. Urai, E. Hotta, M. Maeyama, H. Yasui and T. Tamagawa, "Effect of axial magnetic fields on electrical characteristics of low pressure wire discharge," *Jpn. J. Appl. Phys.*, Vol. 33, pp. 4243-4246, 1994.
3. Y. Nakagawa and H. Kawauchi, "Pulse Intense Electron Beam Irradiation on the Atmospheric Pressure $N_2$ Containing 200 ppm of NO", *Jpn. J. Appl. Phys.,* Vol. 37, pp. 5082-5087, 1998.
4. B. M. Penetrante, J. N. Bardsley and M. C. Hsiao, "Kinetic analysis of non-thermal plasmas used for pollution control," *Jpn. J. Appl. Phys.*, Vol. 36, pp. 5007-5017, 1997.
5. B. M. Penetrante, M. C. Hsiao, B. T. Merritt, G. E. Vogtlin, P. H. Wallman, "Comparison of electrical discharge techniques for non-thermal plasma processing of NO in $N_2$," *IEEE Trans. Plasma Sci.,* Vol. 23, pp. 679-687, 1995.